# Superconducting magnesium diboride films on Silicon with $T_{c0}$ ~ 24K grown via vacuum annealing from stoichiometric precursors

H.Y. Zhai,[*] H.M. Christen, L. Zhang, C. Cantoni, M. Paranthaman, B.C. Sales, D.K. Christen, and D.H. Lowndes

*Oak Ridge National Laboratory, Oak Ridge, TN 37931-6056*

Superconducting magnesium diboride films with $T_{c0}$ ~ 24 K and sharp transition ~ 1 K were successfully prepared on silicon substrates by pulsed laser deposition from stoichiometric $MgB_2$ target. Contrary to previous reports, anneals at 630 °C and a background of $2\times10^{-4}$ torr Ar/4%$H_2$ were performed without the requirement of Mg vapor or an Mg cap layer. This integration of superconducting $MgB_2$ films on silicon may thus prove enabling in superconductor-semiconductor device applications. Images of surface morphology and cross-section profiles by scanning electron microscopy (SEM) show that the films have a uniform surface morphology and thickness. Energy dispersive spectroscopy (EDS) reveals these films were contaminated with oxygen, originating either from the growth environment or from sample exposure to air. The oxygen contamination may account for the low $T_c$ for those *in-situ* annealed films, while the use of Si as the substrate does not result in a decrease in $T_c$ as compared to other substrates.

---

[*] Electronic Mail: zhai@solid.ssd.ornl.gov

The discovery of superconductivity in the binary compound $MgB_2$ is remarkable considering its simple composition, the easily obtained starting materials, and the relatively high $T_c$ among the classic superconductors.[1-2] Recent attention has been focused on physical properties and on potential large-scale applications and devices.[3-5] Only weeks after the discovery, the first superconducting $MgB_2$ thin films were successfully fabricated.[6-10] However, these efforts were complicated by the low decomposition temperature (~ 650 °C) and the high equilibrium phase formation temperature (>800 °C) of $MgB_2$. To date, all $MgB_2$ films with bulk-like $T_c$ > 38 K were prepared with an *ex-situ* method starting from boron films, followed by a post-deposition annealing at high temperature (~ 900 °C) in the presence of Mg in a sealed container.[7,9-10]

Silicon is an attractive substrate for the growth of superconducting films due to the potential for device integration. Attempts to integrate Si and high-$T_c$ oxide superconductors have failed, because Si-contamination into those oxides depresses superconductivity. For $MgB_2$ on Si, post-deposition annealing at high temperature ~ 900 °C, using the same procedure as with other substrates,[7] has not yielded superconducting material, and the only reported $MgB_2$ films grown on Si using *ex-situ* annealing show a low $T_c$ and a wide transition.[11]

This letter reports the systematic study of $MgB_2$ films on silicon substrates by pulsed laser deposition (PLD). The measured $T_{c0}$ is 24 K with a narrow transition width of about 1 K, which is comparable to $MgB_2$ films prepared on other substrates ($LaAlO_3$, $SrTiO_3$, and $Al_2O_3$) using an *in-situ* annealing procedure on Mg-rich precursor films, and either an Mg cap layer[6,8] or an Mg plasma.[7] The present technique for fabricating superconducting $MgB_2$ on Si differs remarkably from those previous reports: (1) The starting precursor films of $MgB_2$ were prepared from a $MgB_2$ target; (2) no excess of Mg or $MgB_2$ was present during the annealing; (3) the films were annealed in $2 \times 10^{-4}$ torr of $Ar/4\%H_2$ (with a base pressure of $5 \times 10^{-6}$). Resistance versus temperature curves were measured using a standard four-probe method in a small refrigerator system capable of reaching 11 K. X-ray diffraction θ-2θ scans were used to study the structure of the $MgB_2$ films. Surface morphologies and cross-section profiles of $MgB_2$ films reacted at different annealing temperatures were examined by scanning electron microscopy (SEM). Film compositions at different sites (surface, middle of the film, $MgB_2$/Si interface) were studied by energy-dispersive spectrometry (EDS). Details of growth mechanisms and composition for $MgB_2$ films fabricated with different processes and substrates will be reported elsewhere.[12]

Stoichiometric targets were prepared using $MgB_2$ powders obtained either from Alfa Aesar or by reacting elemental Mg turnings with amorphous B powder. Both conventionally sintered and hot-pressed pellets were used. Deposition was carried out at room temperature in a background of $10^{-4}$ Torr of $Ar/4\%H_2$. Laser energies ranged from 200 – 400 mJ per pulse at a wavelength 248 nm, corresponding to an energy density of 1.7 – 3.3 $J/cm^2$ on the target, and the laser repetition rate was 15 Hz. Typical thickness of these films was 600 nm. N-type, (100)-oriented Si with a resistivity of 0.2-0.3 Ω·cm was used as the substrate material for all samples described here. Details regarding the experimental system and the target preparation can be found elsewhere.[6,8] In order to start from exactly the same condition, all 6 samples annealed at different temperatures were cut from a single $MgB_2$-coated Si chip measuring about 8 mm by 11 mm. Five of these samples (1-2 mm by 8 mm) were then annealed in the same laser deposition chamber at 500 °C, 550 °C, 600 °C, 630 °C, and 670 °C respectively, with the remaining one serving as a reference. Each anneal was performed with a temperature ramp of 100 °C/min in a $2 \times 10^{-4}$ $Ar/4\%H_2$ background, and the sample was held at the annealing temperature for 20 minutes. The chamber was then vented with 1 atm $Ar/4\%H_2$ and cooled to below 200 °C at 50 °C/min. Considering the relatively slow decomposition of $MgB_2$ even at elevated temperature,[14] these rates are sufficiently high to exclude effects related to the ramp time in our study.

With the current growth conditions, a stoichiometric $MgB_2$ precursor cannot be obtained above 300 °C, and the superconducting $MgB_2$ phase is not formed at room temperature.[8] Plotted in Fig. 1 are the resistance versus temperature curves of $MgB_2$ films grown on Si from the stoichiometric target at room temperature and annealed at 500 °C (Fig. 1a), 550 °C (Fig. 1b), 600 °C (Fig. 1c), 630 °C (Fig. 1d), and 670 °C (Fig. 1e). The film grown at room temperature (not shown) has a flat R-T curve and shows no sign of a transition above 11 K. The resistance of samples annealed above 500 °C increases when the temperature is decreased, presumably due to carrier freeze-out in the Si substrate. Zero resistance is not observed on films annealed at 500 °C, 550 °C, and 670 °C. The film annealed at 600 °C has a $T_{c0}$ > 18 K with $T_c^{onset}$ ~ 25 K, and the film annealed at 630 °C has a $T_{c0}$ at 24 K and $T_c^{onset}$ at 25.5 K. It is particularly interesting to note that Mg-rich precursors on $Al_2O_3$ with Mg cap layers annealed in 1 atm. of $Ar/4\%H_2$ do not yield superconducting films when annealed above 630 °C,[8] whereas 630 °C is the optimum temperature for the current films. (Note that the system used for this work is the same as that used in references 6 and 8.)

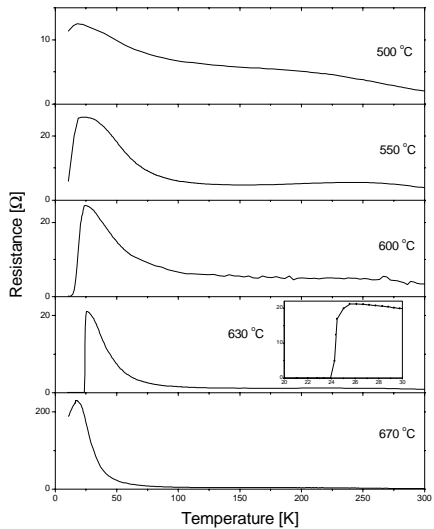

Fig. 1. Resistance – Temperature curves for $MgB_2$ films prepared on n-Si (100) using standard four-probe method. Measurements down to 11 K were performed. Measured resistivity of original Si substrate at 25 K is ~ 300 $\Omega\cdot$cm corresponding to 2 k$\Omega$ as the same size substrate we used in our samples.

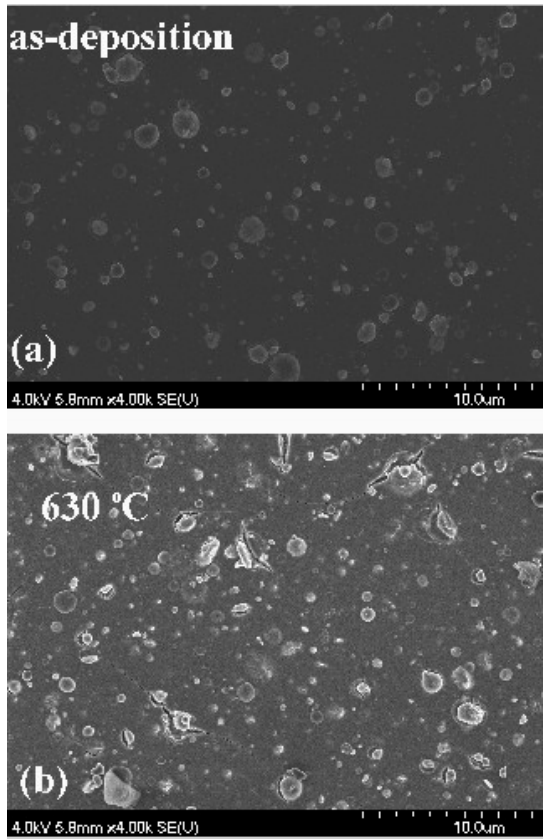

Figure 2. SEM images of surface morphologies of $MgB_2$ films grown at room temperature (a) and annealed at 630 °C (b).

Figure 2 shows SEM images of $MgB_2$ film surface morphologies for different treatments. The $MgB_2$ film deposited at room temperature without high temperature treatment shows a uniform layer with typical micron-size droplets from PLD, as shown in Fig. 2a. The $MgB_2$ films annealed at 500 °C, 550 °C, 600 °C, 630 °C, and 670 °C all show a uniform and identical surface morphology except for the typical micron-size PLD particles, as shown in Fig. 2b for the sample annealed at 630 °C. We also observed cracks in these films and the films grown on other substrates ($Al_2O_3$ etc.) using the *in situ* procedure.[8] The cracks usually originate at a droplet or encircle a droplet, possibly implying chemical composition variations in that area in addition to the different thermal expansion coefficients between the substrate (Si) and $MgB_2$ films.

For the sample annealed at 630 °C ($T_{c0}$ ~ 24K), an SEM cross-section image is shown in Fig. 3. The film is dense with a uniform thickness of 600 nm. EDS performed at different sites show oxygen contamination throughout the film. Because EDS is insensitive to light elements, the B percentage can only be determined semi-quantitatively. Nevertheless, the Mg:B ratio in the film clearly deviates from the bulk value of 0.5, measuring about 0.3 throughout most of the volume and decreasing slightly near the surface. No anomalous Mg-depletion near the film/substrate interface is observed. However,

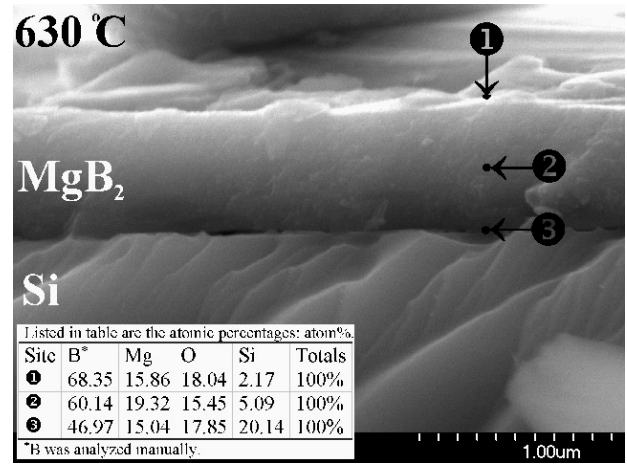

Figure 3. Cross-section SEM images and EDS composition analysis results (inset) on the film annealed at 630 °C.

Mg diffuses substantially into the Si substrate, consistent with the reported formation of an Mg-Si compound at 650 °C.[13] Note that these results may be numerically accurate because EDS typically samples a volume comparable to the present film thickness.

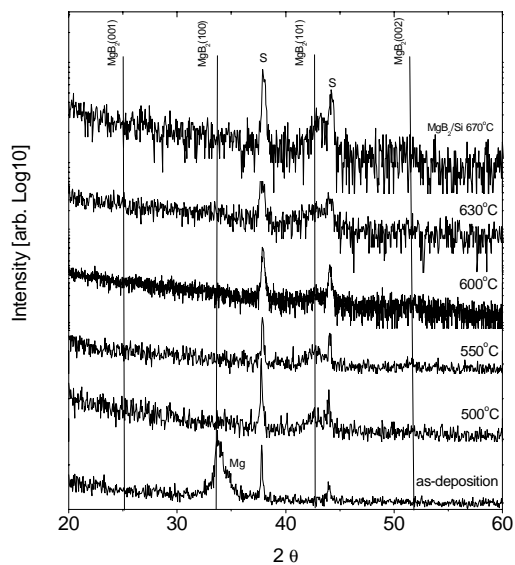

Figure 4. X-ray diffraction θ-2θ scans showing the broad and weak peaks of MgB$_2$ (101) and (002) in films annealed above 500 °C, and the presence of elemental Mg in the as-deposited film.

The films' structure was also studied using X-ray diffraction (Figure 4). As expected from powder diffractograms of polycrystalline MgB$_2$,[7] the MgB$_2$ (101) reflection is strongest for annealed films, with the (002) peak also being detected. Considering the film thickness of about 600 nm, the broad and weak peaks imply that only a small volume fraction of the layers is crystalline MgB$_2$, as was observed previously for Mg-rich MgB$_2$ films annealed *in-situ* with an Mg cap layer.[6,8] Moreover, this result also explains why the resistance of these film is much larger than that of stoichiometric MgB$_2$ materials (Fig. 1). Magnesium deficiency and oxygen contamination are the most likely reasons for the incomplete crystallization, and the oxygen contamination could be caused by oxygen in the growth chamber or in the target materials, or could result from the exposure of the precursor film to air before the anneal. However, films on other substrates, for which the precursor was not exposed to air, showed comparable $T_c$, indicating that exposure to air may not be the dominating factor. Oxygen contamination has also been identified in a transmission electron microscopy study by Eom *et a.*, and the laser induced MgB$_4$/MgB$_7$ formation has been proposed as possible contamination in PLD-grown MgB$_2$ films.[8,15]

In summary, several issues have been identified as limiting factors in the growth of MgB$_2$ films on Si substrates. In particular, the formation of a binary Mg-Si compound at the film/substrate interface restricts the current approach to annealing temperatures below 650 °C, and our data indicate that the Mg+B precursor films do not fully react to form the correct MgB$_2$ phase in anneals at 630 °C. Oxygen contamination and Mg-deficiency are identified as possible reasons for incomplete crystallization, leading to a further decrease in the transition temperatures. Nevertheless, these MgB$_2$ films prepared on n-type, (100)-oriented Si substrates show $T_c \approx 24$ K and sharp transitions with widths of about 1 K. Comparison to films grown on other substrates shows that Si does not suppress the superconducting when annealed at 630 °C. This approach does not rely on an additional Mg vapor source or Mg cap layers for the anneals, contrary to earlier studies, and will thus facilitate the integration of MgB$_2$ layers into heterostructures, possibly enabling the formation of junctions and devices.